\documentclass[11pt,a4paper]{article}

\usepackage{amsmath,amssymb,amsthm}
\usepackage{graphicx}
\usepackage[hidelinks]{hyperref}
\usepackage{cite}
\usepackage{geometry}
\usepackage{setspace}
\usepackage{slashed}
\usepackage{bm}
\usepackage[T1]{fontenc}

\hypersetup{
  colorlinks=true,
  linkcolor=blue,
  citecolor=blue,
  urlcolor=blue
}

\title{\textbf{Sterile Neutrinos as a Dynamical Cosmological Fluid:\\
Implications for the Expansion History and\\
Matter--Radiation Equality}}

\author{
  Poulastya Kar\\
  \small Kirori Mal College, University of Delhi, Delhi, India
  \and
  Bipin Singh Koranga\\
  \small Associate Professor, Kirori Mal College, University of Delhi, Delhi, India
}

\date{}

\begin{document}

\maketitle

\begin{abstract}
Sterile neutrinos appear naturally in many extensions of the Standard
Model and may influence cosmological evolution even when their abundance
remains subdominant. Their cosmological impact is commonly parameterized
by a constant shift in the effective number of relativistic species,
$\Delta N_\mathrm{eff}$, implicitly assuming a radiation-like equation
of state. However, sterile neutrinos with finite mass and incomplete
thermalization generally exhibit a time-dependent equation of state as
the Universe cools and the population transitions from relativistic to
non-relativistic behavior.

In this work we develop an analytic framework that treats sterile
neutrinos as a dynamical cosmological fluid with an evolving equation of
state. Starting from the Boltzmann equation in an expanding
Friedmann--Lema\^{i}tre--Robertson--Walker (FLRW) background, we show
that incomplete thermalization---arising naturally from the suppressed
active-sterile oscillation probability in the primordial
plasma---leads to a suppressed Fermi--Dirac distribution characterized
by a thermalization parameter $\alpha \leq 1$. Using this distribution
we compute the sterile neutrino energy density and pressure and
incorporate them self-consistently into the Friedmann equations.

We demonstrate that partially thermalized sterile neutrinos generically
induce time-dependent modifications to the Hubble expansion rate,
exhibiting three distinct dynamical regimes: a relativistic plateau, a
relativistic--non-relativistic transition, and a matter-like growth
phase. For GeV-scale sterile neutrinos the relativistic fraction at
matter--radiation equality is negligibly small, implying that their
contribution at equality is effectively matter-like rather than
radiation-like. As a consequence, the presence of a sterile component
shifts the equality epoch by an amount proportional to the sterile
energy fraction. Combining this analytic result with observational
constraints on the equality scale yields a bound on the sterile energy
fraction at equality, $f_s \lesssim \mathcal{O}(10^{-2})$. The
framework provides a transparent connection between microscopic sterile
neutrino production physics and macroscopic cosmological expansion, and
highlights a phenomenological regime not captured by the conventional
$\Delta N_\mathrm{eff}$ parameterization.
\end{abstract}

\tableofcontents
\newpage

\section{Introduction}
\label{sec:intro}

The discovery of nonzero neutrino masses provides compelling evidence
for physics beyond the Standard Model~\cite{Minkowski1977,Yanagida1979,GellMann1979,Glashow1980,Mohapatra1980}.
Among the simplest and most widely studied extensions are sterile
neutrinos---gauge-singlet fermions that mix with the active neutrino
sector through Yukawa couplings. Such states arise naturally in a
variety of theoretical frameworks including Type~I seesaw models of
neutrino mass generation~\cite{Minkowski1977,Yanagida1979,GellMann1979},
low-scale leptogenesis scenarios, and hidden-sector
constructions~\cite{Kusenko2009,Abazajian2017}.

Sterile neutrinos are also of considerable interest in
cosmology~\cite{Kusenko2009,Abazajian2017,Abazajian2012}. Even a
relatively small population can modify the thermal history of the early
Universe by contributing to the total energy density and thereby
altering the cosmic expansion rate. The cosmological constraints on
sterile neutrino parameters depend sensitively on the production
mechanism. In the Dodelson--Widrow (DW) non-resonant
mechanism~\cite{Dodelson1994}, sterile neutrinos are produced through
active-sterile oscillations in the early Universe plasma, yielding a
suppressed abundance determined by the vacuum mixing angle. In the
resonant Shi--Fuller mechanism~\cite{Shi1999}, a finite lepton
asymmetry enhances production via the MSW resonance, allowing a larger
sterile abundance at smaller mixing angles. Complementary constraints
arise from X-ray observations searching for the radiative decay
$\nu_s \to \nu_a + \gamma$~\cite{Boyarsky2009,Abazajian2001}, and from
short-baseline oscillation experiments probing eV-scale sterile
states~\cite{Giunti2011,Palazzo2011}.

In many analyses the cosmological effects of additional neutrino species
are parameterized through a shift in the effective number of
relativistic degrees of freedom, $\Delta
N_\mathrm{eff}$~\cite{Lesgourgues2006,Lesgourgues2014,Planck2018}.
This approach is appropriate when the additional species remain fully
relativistic over the cosmological epochs of interest. However, sterile
neutrinos with finite mass generally undergo a transition from
relativistic to non-relativistic behavior as the Universe expands and
cools. In such cases their equation of state evolves dynamically and the
assumption of a constant radiation-like contribution becomes inadequate.

A further complication arises when sterile neutrinos are only partially
thermalized in the early Universe. In many theoretically motivated
scenarios the interaction rate of sterile neutrinos never becomes large
enough for full thermal equilibrium to be established~\cite{Dodelson1994,Shi1999,Kusenko2009}.
The resulting phase-space distribution is therefore suppressed relative
to the equilibrium Fermi--Dirac form, leading to a population that is
both subdominant and dynamically evolving.

\subsection{Previous Work and Motivation}

Previous work on sterile neutrinos in cosmology has largely proceeded
along two distinct tracks. The first treats sterile neutrinos as fully
thermalized species, constraining them via their contribution to
$\Delta N_\mathrm{eff}$ using precision CMB
data~\cite{Lesgourgues2014,Planck2018} and Big Bang
nucleosynthesis~\cite{Hannestad2010,Dolgov2002}. The second focuses on
specific mass windows: keV-scale warm dark matter candidates produced
via the Dodelson--Widrow or Shi--Fuller
mechanisms~\cite{Dodelson1994,Shi1999}, or eV-scale states motivated by
short-baseline oscillation anomalies~\cite{Giunti2011,Palazzo2011}. In
both cases the sterile component is typically characterized by a single
effective parameter---either a contribution to $\Delta N_\mathrm{eff}$
or a mixing angle---rather than treated as a dynamical fluid.

The intermediate regime of partial thermalization with GeV-scale masses,
where the equation of state evolves continuously from radiation-like to
matter-like, has not been treated as a self-consistent cosmological
fluid in the existing literature. Boltzmann codes such as
\textsc{Class}~\cite{Lesgourgues2011,Lesgourgues2011b} and
\textsc{Camb}~\cite{Lewis2000} handle massive neutrinos numerically by
evolving the perturbation hierarchy, but do not provide transparent
analytic insight into the background-level dynamics for partially
thermalized populations. The fluid approximations developed for active
massive neutrinos by Lesgourgues and
Pastor~\cite{Lesgourgues2006,LesgourguesMangano2013} assume thermal
equilibrium distributions and are not directly applicable to the
suppressed Fermi--Dirac form relevant for incomplete thermalization.

Recent studies within the four-flavor neutrino framework have analyzed
the structure of the effective Majorana neutrino mass matrix relevant
for neutrinoless double-beta decay, demonstrating significant
sensitivity to additional mixing angles and CP-violating
phases~\cite{Koranga2021a}. The behavior of CP violation through
modifications of the Jarlskog invariant near the grand unification scale
has also been examined, highlighting the potential impact of new physics
on leptonic CP observables~\cite{Koranga2021b}. Together these studies
provide a phenomenological basis for understanding how sterile neutrinos
may influence neutrino mass generation and mixing properties, motivating
the present fluid-level treatment.

While numerical Boltzmann codes such as \textsc{Class} and \textsc{Camb}
evolve the full perturbation hierarchy for massive neutrinos, they do
not provide closed-form expressions for the background-level dynamics of
partially thermalized populations. The present framework fills this gap:
the analytic parametrization in terms of $a_\mathrm{nr}$ and $\alpha$
offers a direct and transparent mapping between microphysical production
parameters and the cosmological expansion history, without requiring
numerical integration of the full kinetic equations.

\subsection{This Work}

These features motivate a more general framework for describing sterile
neutrino cosmology beyond the conventional $\Delta N_\mathrm{eff}$
parameterization. In particular, a realistic treatment should
incorporate both the finite mass of the sterile neutrino and the
possibility of incomplete thermalization. Together these effects imply
that the sterile neutrino population behaves as a dynamical cosmological
fluid whose equation of state evolves continuously from radiation-like
to matter-like~\cite{Baumann2018,Kolb1990}.

In this work we develop an analytic framework for describing sterile
neutrinos in precisely this regime. Starting from the Boltzmann equation
in an expanding FLRW spacetime~\cite{Weinberg2008,Dodelson2003,Baumann2018,Kolb1990},
we derive the parametric conditions under which incomplete thermalization
leads to a suppressed Fermi--Dirac distribution. We then compute the
corresponding sterile neutrino energy density and pressure and
incorporate them directly into the Friedmann equations governing the
cosmological background evolution.

A key result of our analysis is that partially thermalized sterile
neutrinos generically induce time-dependent modifications to the
expansion rate of the Universe, exhibiting three distinct dynamical
regimes. These effects are particularly relevant near the
relativistic--non-relativistic transition, where the sterile equation of
state departs most strongly from that of standard radiation. As a
concrete application we study the impact of sterile neutrinos on the
epoch of matter--radiation equality, whose timing is tightly constrained
by CMB observations~\cite{Planck2018}.

For sterile neutrinos in the GeV mass range we show that the
relativistic fraction at equality is negligibly small. The sterile
population therefore contributes effectively as matter at that epoch,
leading to a shift in the equality scale that depends directly on the
sterile energy fraction rather than on a radiation-like correction. This
provides a simple analytic connection between sterile neutrino cosmology
and observational constraints on the expansion history.

The key novelty of this work is threefold: (i)~we formulate sterile
neutrinos with incomplete thermalization as a dynamical cosmological
fluid with a time-dependent equation of state, going beyond the constant
$\Delta N_\mathrm{eff}$ description; (ii)~we identify three distinct
dynamical regimes of the expansion-rate correction and provide analytic
estimates for each; and (iii)~we show that GeV-scale sterile neutrinos
contribute as matter---not radiation---at equality, yielding a
constraint on $f_s$ that is qualitatively distinct from $\Delta
N_\mathrm{eff}$ bounds.

The structure of the paper is as follows. In Sec.~\ref{sec:fluid} we
formulate sterile neutrinos as a dynamical cosmological fluid and derive
the microscopic origin of incomplete thermalization. In
Sec.~\ref{sec:friedmann} we incorporate the sterile component into the
Friedmann equations and estimate its impact on the cosmic expansion
rate. In Sec.~\ref{sec:equality} we derive analytic constraints from the
stability of the matter--radiation equality epoch. Numerical
illustrations are presented in Sec.~\ref{sec:numerical}. Limitations of
the present framework are discussed in Sec.~\ref{sec:limitations}. We
discuss our results and outline future directions in
Sec.~\ref{sec:discussion}.

\section{Sterile Neutrinos as a Dynamical Cosmological Fluid}
\label{sec:fluid}

We begin by formulating the sterile neutrino population as a dynamical
component of the cosmological energy budget. We consider a spatially
flat FLRW spacetime~\cite{Weinberg2008,Baumann2018} with metric
\begin{equation}
  ds^2 = -dt^2 + a^2(t)\,d\mathbf{x}^2,
  \label{eq:metric}
\end{equation}
where $a(t)$ is the cosmological scale factor. The Universe is assumed
to be described by standard General Relativity, with no modification of
the gravitational sector. The total cosmic energy density contains
contributions from radiation, matter, and a sterile neutrino component.

We introduce a sterile neutrino species $\nu_s$ with mass $m_s$,
assumed to be a gauge-singlet fermion that interacts with the Standard
Model through mixing with active neutrinos. After decoupling from the
primordial plasma, the sterile neutrino population evolves
collisionlessly and is described by a phase-space distribution function
$f_s(p,t)$. It is convenient to express the distribution as a function
of the comoving momentum $q = ap$, which remains constant during free
streaming~\cite{Dodelson2003,Kolb1990}.

\subsection{Seesaw Lagrangian and the Microscopic Basis for Mixing}
\label{sec:seesaw}

The sterile states $N_I$ are introduced as gauge-singlet Majorana
fermions. In the Type~I seesaw
framework~\cite{Minkowski1977,Yanagida1979,GellMann1979,Glashow1980,Mohapatra1980},
the most general renormalizable Lagrangian extending the Standard Model
is
\begin{equation}
  \mathcal{L} = \mathcal{L}_\mathrm{SM}
    + i\,\bar{N}_I \gamma^\mu \partial_\mu N_I
    - \left(Y_{\alpha I}\,\bar{L}_\alpha \tilde{\Phi}\, N_I
      + \tfrac{1}{2} M_I\,\bar{N}^c_I N_I + \mathrm{h.c.}\right),
  \label{eq:seesaw_lag}
\end{equation}
where $L_\alpha$ is the lepton doublet, $\tilde{\Phi}$ is the conjugate
Higgs doublet, $Y_{\alpha I}$ are the Yukawa couplings, and $M_I$ is
the Majorana mass scale~\cite{Minkowski1977,Yanagida1979,GellMann1979}.
After electroweak symmetry breaking, the Higgs acquires a vacuum
expectation value $v/\sqrt{2}$, generating the Dirac mass matrix $M_D =
Y v/\sqrt{2}$. The active-sterile mixing is then governed by the matrix
\begin{equation}
  U \approx M_D M^{-1},
  \label{eq:mixing_matrix}
\end{equation}
and the light neutrino mass is suppressed as
\begin{equation}
  m_\nu \sim M_D^2 M^{-1} \sim U^2 M.
  \label{eq:seesaw}
\end{equation}
This provides the microscopic basis for the mixing angle $\theta$ used
throughout our fluid description: the vacuum mixing angle $\sin^2\theta
\sim |U|^2$ relates the flavor eigenstate to the mass eigenstate, and
its smallness is a direct consequence of the large Majorana mass scale
$M_I \gg M_D$.

The active-sterile mixing angle $\theta$ plays a dual role in
cosmology: it controls both the production rate of sterile neutrinos in
the early Universe and the decay rate through the channel $\nu_s \to
\nu_a + \gamma$. Observational bounds on the latter from X-ray
observations~\cite{Boyarsky2009,Abazajian2001} constrain the
$(m_s,\sin^2\theta)$ parameter space from above. Within the remaining
window, the condition of incomplete thermalization defines an upper
boundary via the full thermalization condition and a lower boundary via
the BBN decay constraint, as discussed in
Sec.~\ref{sec:param_space} and illustrated in Fig.~\ref{fig:param_space}.

\subsection{Sterile Neutrino Distribution Function}
\label{sec:distrib}

In many theoretically motivated scenarios sterile neutrinos never reach
full thermal equilibrium in the early
Universe~\cite{Dodelson1994,Shi1999,Kusenko2009,Abazajian2017}. Instead
their production rate remains comparable to or smaller than the Hubble
expansion rate, resulting in incomplete thermalization. In this regime
the sterile neutrino distribution can be approximated by a suppressed
Fermi--Dirac form~\cite{Dodelson1994,Dolgov2002}
\begin{equation}
  f_s(p) = \frac{\alpha}{e^{p/T_s}+1},
  \label{eq:fd}
\end{equation}
where $T_s$ is an effective sterile neutrino temperature and $0 \leq
\alpha \leq 1$ parameterizes the degree of thermalization. The case
$\alpha = 1$ corresponds to full thermal equilibrium, while $\alpha \ll
1$ describes a strongly suppressed sterile population. After decoupling,
the sterile neutrino temperature redshifts as $T_s \propto
a^{-1}$~\cite{Dodelson2003,Kolb1990}.

\subsection{Energy Density and Pressure}
\label{sec:rho_p}

The macroscopic effect of the sterile neutrino population on
cosmological expansion is determined by its energy-momentum
tensor~\cite{Weinberg2008,LesgourguesMangano2013,Kolb1990}. For an
isotropic distribution function the sterile neutrino component behaves
as a perfect fluid with energy-momentum tensor $T^\mu{}_\nu =
\mathrm{diag}(-\rho_s, P_s, P_s, P_s)$. The energy density and
pressure are obtained by integrating over phase
space~\cite{LesgourguesMangano2013,Kolb1990}
\begin{align}
  \rho_s(a) &= \frac{g_s}{2\pi^2}
    \int_0^\infty dp\, p^2 \sqrt{p^2+m_s^2}\; f_s(p,a),
  \label{eq:rho_s} \\[4pt]
  P_s(a) &= \frac{g_s}{6\pi^2}
    \int_0^\infty dp\, \frac{p^4}{\sqrt{p^2+m_s^2}}\; f_s(p,a),
  \label{eq:P_s}
\end{align}
where $g_s$ denotes the number of internal degrees of freedom. For a
Majorana sterile neutrino $g_s = 2$, counting the two spin states;
since a Majorana fermion is its own antiparticle, no additional factor
arises from particle-antiparticle doubling~\cite{Kolb1990}. Expressing
these in terms of the comoving momentum $q =
ap$~\cite{Dodelson2003,LesgourguesMangano2013}
\begin{align}
  \rho_s(a) &= \frac{g_s}{2\pi^2 a^4}
    \int_0^\infty dq\, q^2 \sqrt{q^2+(am_s)^2}\; f_s(q),
  \label{eq:rho_s_q} \\[4pt]
  P_s(a) &= \frac{g_s}{6\pi^2 a^4}
    \int_0^\infty dq\, \frac{q^4}{\sqrt{q^2+(am_s)^2}}\; f_s(q).
  \label{eq:P_s_q}
\end{align}
These expressions make explicit the dependence of the sterile neutrino
energy density on the scale factor.

\subsection{Time-Dependent Equation of State}
\label{sec:eos}

A useful quantity characterizing the sterile neutrino fluid is the
equation-of-state parameter~\cite{Weinberg2008,Dodelson2003}
\begin{equation}
  w_s(a) = \frac{P_s(a)}{\rho_s(a)}.
  \label{eq:eos}
\end{equation}
Because the sterile neutrino population contains particles with finite
mass, the equation of state evolves with the expansion of the Universe.
In the relativistic regime $p \gg m_s$, the energy-momentum relation
reduces to $E \approx p$, and the sterile neutrino behaves as radiation
with $w_s \to 1/3$. In the non-relativistic regime $p \ll m_s$, the
particle energy is dominated by the rest mass and the pressure becomes
negligible, so $w_s \to 0$~\cite{Lesgourgues2006,LesgourguesMangano2013}.
The sterile neutrino component thus interpolates continuously between
radiation-like and matter-like behavior as the Universe cools. This
evolving equation of state is the key feature distinguishing sterile
neutrinos with finite mass from additional relativistic species
parameterized by $\Delta
N_\mathrm{eff}$~\cite{Lesgourgues2006,Lesgourgues2014}.

\subsection{Microscopic Origin of Incomplete Thermalization and the Effective Mixing Angle}
\label{sec:micro}

The suppression parameter $\alpha$ introduced above has a precise
microscopic origin in the dynamics governing sterile neutrino production
in the early Universe. Sterile neutrinos couple to the Standard Model
through the Yukawa interactions in Eq.~(\ref{eq:seesaw_lag}). The
phase-space distribution evolves according to the Boltzmann equation in
an expanding Universe~\cite{Dodelson2003,Kolb1990}
\begin{equation}
  \frac{\partial f_s}{\partial t}
  - H p\,\frac{\partial f_s}{\partial p}
  = \Gamma_s(p,T)\bigl(f_s^\mathrm{eq} - f_s\bigr),
  \label{eq:boltzmann}
\end{equation}
where $H$ is the Hubble expansion rate and $\Gamma_s$ is the
interaction rate responsible for sterile neutrino production. During
radiation domination the Hubble rate is
approximately~\cite{Baumann2018,Kolb1990}
\begin{equation}
  H(T) = 1.66\,\sqrt{g_*}\,\frac{T^2}{M_\mathrm{Pl}}.
  \label{eq:hubble_rad}
\end{equation}
The efficiency of sterile neutrino production is controlled by the ratio
$K(T) = \Gamma_s/H$. If $K \gg 1$ the sterile neutrino population
reaches thermal equilibrium; if $K \ll 1$ production is inefficient and
the sterile abundance remains strongly suppressed. In the intermediate
regime $K \lesssim 1$ near the production temperature, the distribution
function can be approximated by a scaled equilibrium form $f_s(p) \simeq
\alpha f_s^\mathrm{eq}(p)$, with $\alpha \sim
(\Gamma_s/H)|_{T\sim T_\mathrm{prod}}$~\cite{Dodelson1994,Dolgov2002}.

Crucially, the production rate $\Gamma_s$ is not governed by the vacuum
mixing angle $\theta$ alone. In the early Universe plasma, the effective
mixing angle is modified by the finite-temperature thermal potential
$V_T$ and a damping rate $\Gamma$ due to active neutrino
scattering~\cite{Shi1999,Dolgov2002}. The effective mixing angle in
matter is given by~\cite{Shi1999,Barbieri1991}
\begin{equation}
  \sin^2 2\theta_\mathrm{eff} \approx
  \frac{\Delta^2 \sin^2 2\theta}
       {\Delta^2 \sin^2 2\theta + (\Gamma/2)^2
        + (\Delta\cos 2\theta - V_T)^2},
  \label{eq:theta_eff}
\end{equation}
where $\Delta = m_s^2/(2E)$ is the vacuum oscillation frequency. At
high temperatures $V_T \gg \Delta\cos 2\theta$, the effective mixing is
strongly suppressed below its vacuum value, so $\theta_\mathrm{eff} \ll
\theta$. As the Universe cools and $V_T$ decreases, $\theta_\mathrm{eff}$
rises toward $\theta$. In the Shi--Fuller resonant mechanism~\cite{Shi1999},
$V_T$ passes through zero at a specific temperature for a nonzero lepton
asymmetry, producing a resonant enhancement of sterile neutrino
production at a narrowly defined epoch.

The parameter $\alpha$ in our dynamical fluid model effectively captures
the time-integrated effect of this suppressed oscillation probability,
representing the fraction of the active neutrino population that has
successfully transitioned into the sterile sector before the interaction
rate $\Gamma$ falls below the Hubble expansion rate $H$. This provides
a direct and physically motivated connection between the microscopic
mixing parameters $(m_s, \sin^2\theta)$ of Fig.~\ref{fig:param_space}
and the macroscopic thermalization factor $\alpha$ appearing in the
fluid description.

\subsection{Parameter Space for Partial Thermalization}
\label{sec:param_space}

\begin{figure}[htbp]
  \centering
  \includegraphics[width=0.82\textwidth]{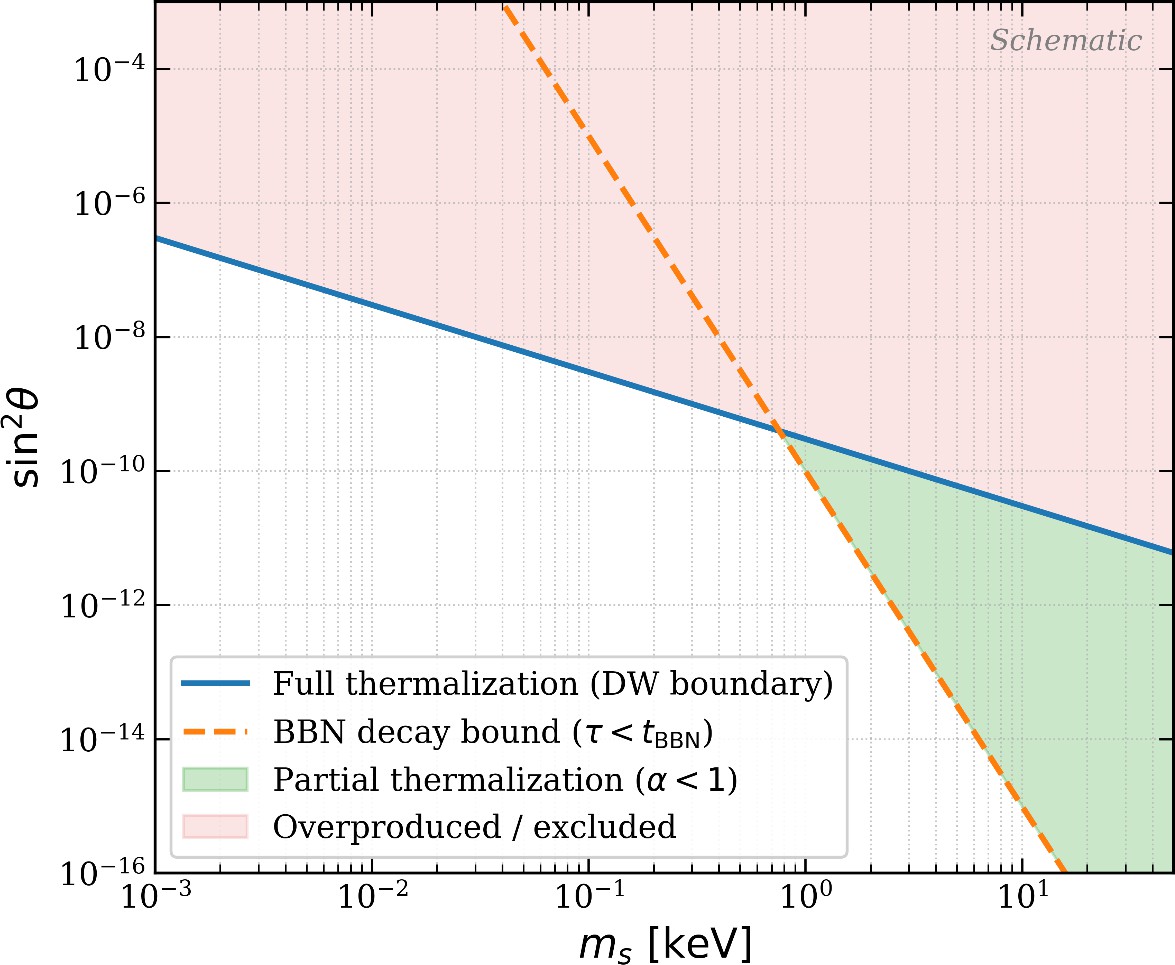}
  \caption{Parameter space for sterile neutrinos in the
    $(m_s,\sin^2\theta)$ plane (schematic). The solid line denotes the
    full thermalization boundary defined by $\Gamma_s/H \sim 1$, with
    $\sin^2\theta \propto m_s^{-1}$ scaling consistent with the
    Dodelson--Widrow production rate~\cite{Dodelson1994}. The dashed
    line corresponds to the requirement of decay before BBN, with
    $\sin^2\theta \propto m_s^{-5}$ from the hadronic decay
    width~\cite{Dolgov2002}. The green shaded region represents the
    regime of partial thermalization ($\alpha < 1$), where sterile
    neutrinos behave as a dynamical cosmological fluid with a
    time-dependent equation of state. The red shaded region denotes the
    overproduced/excluded parameter space above the full thermalization
    boundary.}
  \label{fig:param_space}
\end{figure}

Figure~\ref{fig:param_space} shows the representative parameter region
in the $(m_s, \sin^2\theta)$ plane that corresponds to the regime of
partial thermalization relevant for the dynamical effects discussed in
this work.

The full thermalization boundary (solid line) corresponds to the
condition $K(T_\mathrm{prod}) = \Gamma_s/H \sim 1$. The production rate
for active-sterile oscillation-driven production scales approximately as
$\Gamma_s \sim G_F^2 T^5 \sin^2\theta$~\cite{Dodelson1994,Dolgov2002}.
Setting $\Gamma_s = H(T_\mathrm{prod})$ with $H \sim T^2/M_\mathrm{Pl}$
and evaluating at $T_\mathrm{prod} \sim m_s$ yields
\begin{equation}
  \sin^2\theta\big|_\mathrm{FT} \sim
  \frac{g_*^{1/2}}{M_\mathrm{Pl}\,G_F^2\,m_s^3},
  \label{eq:FT_boundary}
\end{equation}
which decreases with increasing $m_s$. Sterile neutrinos with mixing
angles above this curve are fully thermalized and contribute to $\Delta
N_\mathrm{eff}$ in the standard way; their cosmological effects are
well-constrained by existing
analyses~\cite{Lesgourgues2014,Planck2018,Hannestad2010}.

The BBN decay boundary (dashed line) requires that sterile neutrinos
decay before Big Bang nucleosynthesis, i.e.\ $\tau_s = 1/\Gamma_\mathrm{decay}
< t_\mathrm{BBN}$, where $T_\mathrm{BBN} \sim 1\,\mathrm{MeV}$. The
dominant decay channel has width~\cite{Kolb1990,Dolgov2002}
\begin{equation}
  \Gamma_\mathrm{decay} \sim \frac{G_F^2\,m_s^5\,\sin^2\theta}{192\pi^3},
  \label{eq:decay_width}
\end{equation}
which sets a lower bound on $\sin^2\theta$ at each mass. Sterile
neutrinos with mixing angles below this curve are too long-lived and
would decay during or after BBN, depositing entropy and distorting the
light element abundances~\cite{Hannestad2010,Dolgov2002}.

The shaded region between the two curves defines the
phenomenologically viable window in which sterile neutrinos remain
partially thermalized ($\alpha < 1$) while decaying sufficiently early
to avoid strong cosmological constraints. This is the parameter space in
which the dynamical expansion effects---the time-dependent equation of
state, the three-regime expansion correction, and the matter-like
contribution to equality---described throughout this work are operative.

\section{Modification of the Friedmann Dynamics}
\label{sec:friedmann}

Having established the sterile neutrino fluid description, we now
incorporate this component into the cosmological expansion equations.
Throughout this work we assume a spatially flat FLRW Universe described
by the standard Friedmann equation~\cite{Weinberg2008,Baumann2018}
\begin{equation}
  H^2(a) = \frac{8\pi G}{3}
    \bigl[\rho_r(a) + \rho_m(a) + \rho_s(a)\bigr],
  \label{eq:friedmann}
\end{equation}
where $\rho_r$ denotes the radiation energy density, $\rho_m$ the
matter density, and $\rho_s$ the sterile neutrino contribution. The
presence of a sterile neutrino population therefore modifies the
expansion rate directly through its contribution to the total cosmic
energy density. The evolution of $\rho_s$ is governed by
energy-momentum conservation~\cite{Weinberg2008,Dodelson2003}
\begin{equation}
  \frac{d\rho_s}{da} = -\frac{3}{a}\bigl[1+w_s(a)\bigr]\rho_s(a),
  \label{eq:continuity}
\end{equation}
where $w_s(a) = P_s(a)/\rho_s(a)$ is the equation-of-state parameter
derived in Sec.~\ref{sec:eos}. Because $w_s(a)$ evolves from $1/3$ to
$0$, the sterile neutrino energy density interpolates smoothly between
radiation-like scaling $\rho_s \propto a^{-4}$ and matter-like scaling
$\rho_s \propto a^{-3}$~\cite{Lesgourgues2006,LesgourguesMangano2013}.
This transition introduces a time-dependent modification to the
expansion rate that cannot, in general, be captured by a constant shift
in the relativistic energy density~\cite{Lesgourgues2014,Planck2018}.

\subsection{Parametric Estimate of the Expansion Correction}
\label{sec:param_estimate}

To estimate the magnitude of this effect, consider the relativistic
regime $T \gg m_s$, where the sterile neutrino energy density takes the
form~\cite{LesgourguesMangano2013,Kolb1990}
\begin{equation}
  \rho_s = \alpha\,\frac{7}{8}\,\frac{\pi^2}{30}\,g_s\,T^4.
  \label{eq:rho_rel}
\end{equation}
Here $g_s = 2$ for a Majorana sterile neutrino and $\alpha \leq 1$
characterizes the degree of thermalization. The standard radiation
energy density is $\rho_r = (\pi^2/30)g_* T^4$~\cite{Kolb1990}, where
$g_*$ denotes the effective number of relativistic degrees of freedom.
The fractional change in the Hubble expansion rate induced by the
sterile component is then~\cite{Weinberg2008,Baumann2018}
\begin{equation}
  \frac{\Delta H}{H} = \frac{1}{2}\frac{\rho_s}{\rho_r}
    = \frac{1}{2}\,\frac{7\,g_s\,\alpha}{8\,g_*}.
  \label{eq:DH_H}
\end{equation}
For representative early-Universe values $g_* \simeq 100$ and $g_s =
2$, this yields $\Delta H/H \simeq 0.009\alpha$. Even partial
thermalization can therefore generate measurable corrections to the
cosmic expansion rate, from the percent level ($\alpha = 1$) to the
permille level ($\alpha = 0.1$).

\subsection{Three Dynamical Regimes of the Expansion-Rate Correction}
\label{sec:three_regimes}

\begin{figure}[htbp]
  \centering
  \includegraphics[width=0.82\textwidth]{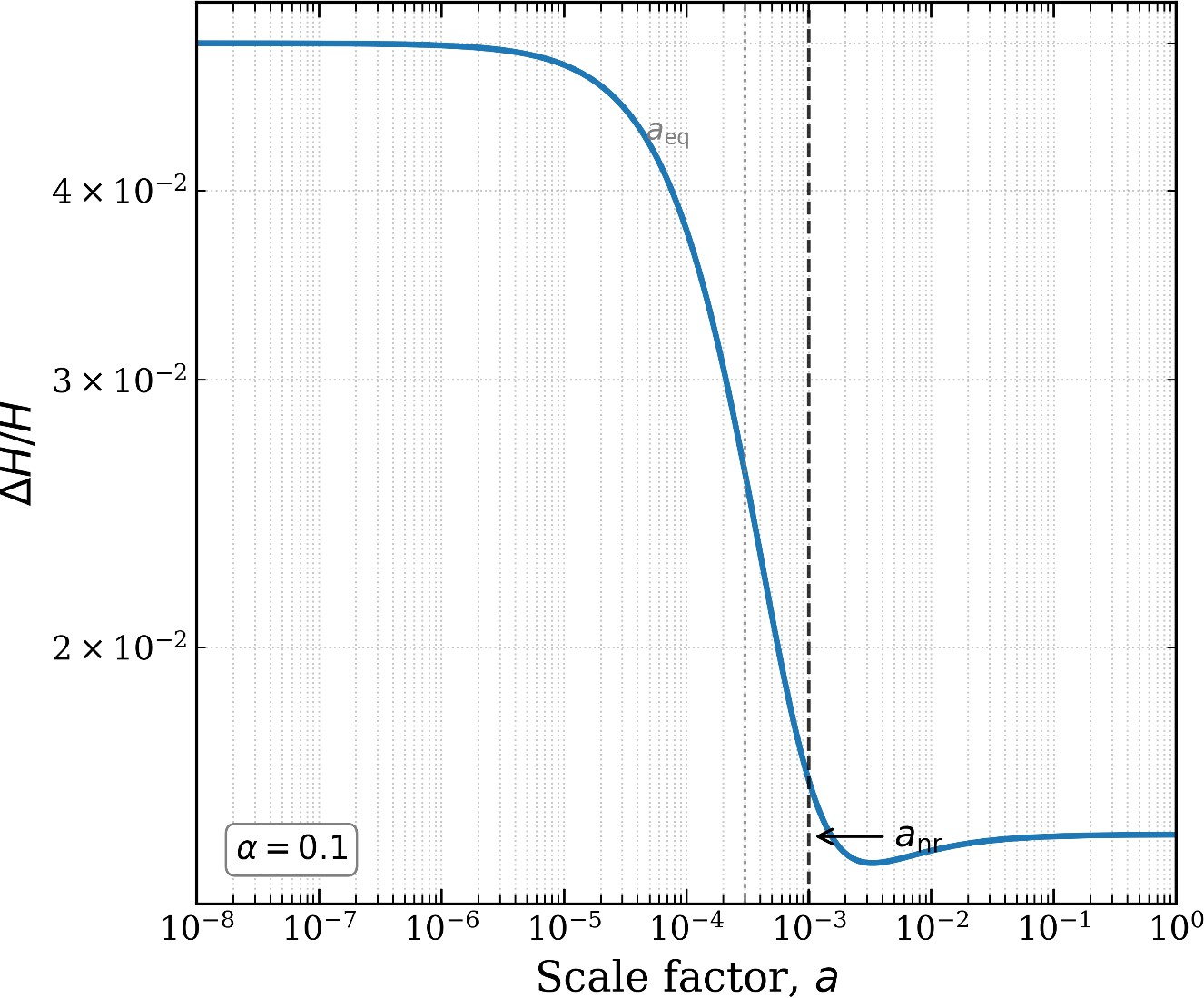}
  \caption{Fractional modification to the Hubble expansion rate,
    $\Delta H/H$, induced by a partially thermalized sterile neutrino
    component ($\alpha = 0.1$). Three distinct regimes are visible:
    (I)~a relativistic plateau at early times where $\Delta H/H \approx
    \mathrm{const}$, (II)~a transition regime near $a \sim a_\mathrm{nr}$
    where the equation of state evolves rapidly, and (III)~a matter-like
    phase where the sterile contribution grows relative to radiation. The
    vertical dashed line marks $a_\mathrm{nr}$ and the dotted vertical
    line marks $a_\mathrm{eq}$. This behavior cannot be captured by a
    constant $\Delta N_\mathrm{eff}$ parameterization.}
  \label{fig:DH}
\end{figure}

The fractional modification to the Hubble expansion rate,
\begin{equation}
  \frac{\Delta H}{H} = \frac{1}{2}\,\frac{\rho_s}{\rho_r + \rho_m},
  \label{eq:DH_H_full}
\end{equation}
exhibits three distinct regimes dictated by the sterile neutrino's
dynamical evolution, illustrated in Fig.~\ref{fig:DH}.

\medskip
\noindent\textbf{Regime~I --- The Relativistic Plateau ($a < a_\mathrm{nr}$):}
At early times, when the sterile neutrino temperature satisfies $T_s \gg
m_s$, the sterile population is deeply relativistic ($p \gg m_s$) and
behaves as a radiation component~\cite{Lesgourgues2006,LesgourguesMangano2013}.
In this regime $\rho_s$ scales as $a^{-4}$, tracking the background
radiation density $\rho_r$. The correction $\Delta H/H$ remains
approximately constant at $0.009\alpha$. This plateau represents the
effective $\Delta N_\mathrm{eff}$ phase of the sterile fluid, in which
the conventional parameterization is a good approximation.

\medskip
\noindent\textbf{Regime~II --- The Relativistic--Non-relativistic
Transition ($a \sim a_\mathrm{nr}$, defined by $T_s \sim m_s$):} As the
Universe cools and the sterile neutrino temperature approaches the
particle mass, the particles begin losing their kinetic energy. The
equation of state $w_s$ begins its descent from $1/3$ toward
$0$~\cite{Lesgourgues2006,LesgourguesMangano2013}. This is the regime
where our dynamical fluid model departs most significantly from standard
$\Delta N_\mathrm{eff}$ approximations, as the assumption of a fixed
radiation-like equation of state breaks down. The largest instantaneous
deviations from standard Friedmann evolution occur at this epoch.

\medskip
\noindent\textbf{Regime~III --- The Matter-like Growth Phase ($a > a_\mathrm{nr}$):}
Once non-relativistic, the sterile energy density scales as $\rho_s
\propto a^{-3}$. Because the background radiation continues to dilute
faster ($\rho_r \propto a^{-4}$), the sterile component's relative
contribution to the total energy density
grows~\cite{Dodelson2003,Kolb1990}. This has a direct consequence for
the equality epoch: the sterile fluid transitions from behaving like
radiation (which delays equality) to behaving like matter (which
advances equality). As the Universe enters matter domination, the
sterile fluid tracks $\rho_m \propto a^{-3}$ and the relative
correction stabilizes.

The features visible in Fig.~\ref{fig:DH} near $a \sim 10^{-2}$
delineate this transition epoch for the GeV-scale mass adopted in our
numerical evaluation.

\section{Constraints from Matter--Radiation Equality}
\label{sec:equality}

The epoch of matter--radiation equality marks the transition between
radiation-dominated and matter-dominated expansion. Its timing plays a
central role in cosmology because it affects the growth of density
perturbations and the acoustic structure of the CMB~\cite{Dodelson2003,Planck2018}.
The equality scale is constrained at the percent level by Planck
observations~\cite{Planck2018}, making it a powerful probe of additional
contributions to the cosmic energy budget.

In standard cosmology equality is defined by the condition
$\rho_m(a_\mathrm{eq}) = \rho_r(a_\mathrm{eq})$~\cite{Dodelson2003,Baumann2018}.
In the presence of a sterile neutrino population the total energy
density receives an additional contribution. Because sterile neutrinos
with finite mass can be partially relativistic near equality, it is
useful to decompose their energy density into radiation-like and
matter-like parts~\cite{Lesgourgues2006,LesgourguesMangano2013}
\begin{equation}
  \rho_s(a) = \rho_s^{(r)}(a) + \rho_s^{(m)}(a),
  \label{eq:decompose}
\end{equation}
where $\rho_s^{(r)} \propto a^{-4}$ represents the relativistic
contribution and $\rho_s^{(m)} \propto a^{-3}$ the matter-like
component. The modified equality condition is
\begin{equation}
  \rho_m(a'_\mathrm{eq}) + \rho_s^{(m)}(a'_\mathrm{eq})
  = \rho_r(a'_\mathrm{eq}) + \rho_s^{(r)}(a'_\mathrm{eq}).
  \label{eq:eq_modified}
\end{equation}

\subsection{Equality Shift}
\label{sec:eq_shift}

Let $a_\mathrm{eq}$ denote the equality scale factor in the standard
cosmological model and $a'_\mathrm{eq}$ the modified value in the
presence of sterile neutrinos. Expanding
Eq.~(\ref{eq:eq_modified}) to leading order in the sterile
contribution~\cite{Dodelson2003,Lesgourgues2006} yields
\begin{equation}
  \frac{\Delta a_\mathrm{eq}}{a_\mathrm{eq}} \simeq f_s(2\eta - 1),
  \label{eq:eq_shift}
\end{equation}
where $f_s = \rho_s/(\rho_r + \rho_m)|_{a_\mathrm{eq}}$ is the sterile
energy fraction at equality and $\eta = \rho_s^{(r)}/\rho_s$ is the
relativistic fraction of the sterile
population~\cite{Lesgourgues2006,LesgourguesMangano2013}. If $\eta
\approx 1$ the sterile population delays equality; if $\eta \approx 0$
it advances equality.

\subsection{Relativistic Fraction at Equality}
\label{sec:rel_frac}

In the relativistic limit, $P_s \simeq \rho_s/3$ implies that the
relativistic fraction $\eta = \rho_s^{(r)}/\rho_s$ approaches unity;
more generally, for a suppressed Fermi--Dirac distribution one finds
$\eta \simeq 3w_s(a)$ as a useful interpolating
approximation~\cite{LesgourguesMangano2013}, which becomes exact in the
ultra-relativistic regime and vanishes with $w_s$ in the
non-relativistic limit. At matter--radiation equality the photon
temperature is $T_\mathrm{eq} \simeq 0.8\,\mathrm{eV}$~\cite{Planck2018,Baumann2018}.
For sterile neutrinos in the GeV mass range, $T_\mathrm{eq}/m_s \sim
10^{-9}$. This hierarchy implies that sterile neutrinos are deeply
non-relativistic at equality, so $w_s(a_\mathrm{eq}) \ll 1$ and
therefore $\eta(a_\mathrm{eq}) \ll
1$~\cite{Lesgourgues2006,LesgourguesMangano2013}. The sterile
population thus behaves effectively as pressureless matter at equality.

\subsection{Constraint on the Sterile Energy Fraction}
\label{sec:constraint}

In the regime $\eta \ll 1$, Eq.~(\ref{eq:eq_shift}) simplifies
to~\cite{Dodelson2003,Planck2018}
\begin{equation}
  \frac{\Delta a_\mathrm{eq}}{a_\mathrm{eq}} \simeq -f_s.
  \label{eq:shift_simple}
\end{equation}
Planck observations constrain the equality scale with approximately
percent-level precision~\cite{Planck2018}. Requiring that the sterile
component does not shift the equality epoch beyond observational limits
therefore implies
\begin{equation}
  \left|\frac{\Delta a_\mathrm{eq}}{a_\mathrm{eq}}\right|
  \lesssim \mathcal{O}(10^{-2}),
  \label{eq:planck_limit}
\end{equation}
which immediately yields the bound
\begin{equation}
  f_s(a_\mathrm{eq}) \lesssim \mathcal{O}(10^{-2}).
  \label{eq:fs_bound}
\end{equation}

\begin{figure}[htbp]
  \centering
  \includegraphics[width=0.85\textwidth]{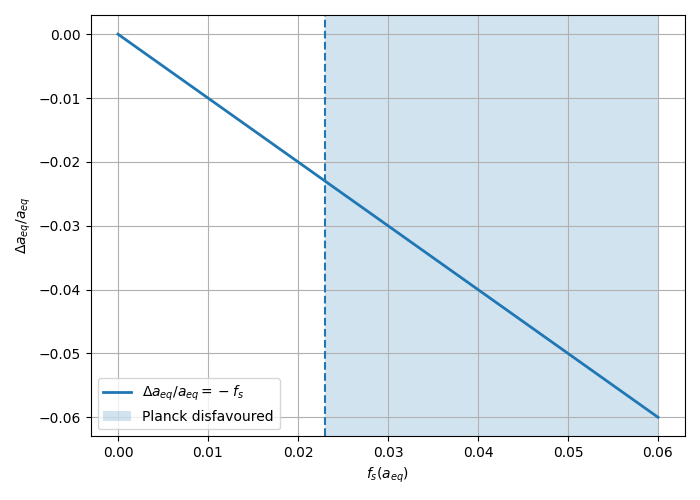}
  \caption{Fractional shift in the matter--radiation equality scale
    factor, $\Delta a_\mathrm{eq}/a_\mathrm{eq}$, as a function of the
    sterile energy fraction $f_s$ at equality. For GeV-scale sterile
    neutrinos the population is deeply non-relativistic at equality,
    implying $\eta \ll 1$ and yielding $\Delta a_\mathrm{eq}/a_\mathrm{eq}
    \simeq -f_s$. The shaded region denotes parameter values disfavoured
    by Planck constraints on the equality scale. Consistency with
    observations requires $f_s(a_\mathrm{eq}) \lesssim
    \mathcal{O}(10^{-2})$. The vertical dashed line marks $f_s = 0.023$,
    corresponding to $\Delta N_\mathrm{eff} = 0.3$ from Planck
    2018~\cite{Planck2018}.}
  \label{fig:eq_shift}
\end{figure}

This constraint differs qualitatively from conventional bounds on
$\Delta N_\mathrm{eff}$~\cite{Lesgourgues2014,Planck2018,Hannestad2010}.
Because the sterile population is already non-relativistic at equality,
its cosmological signature arises through a matter-like contribution to
the energy density rather than through additional radiation. The
equality scale therefore provides a complementary probe of sterile
neutrino cosmology that is sensitive to the dynamical equation of state
of the sterile population~\cite{Lesgourgues2006,LesgourguesMangano2013}.

\subsection{Free-Streaming Scale of Sterile Neutrinos}
\label{sec:free_stream}

In addition to modifying the background expansion rate, massive
neutrinos can influence cosmological structure formation through free
streaming~\cite{Lesgourgues2006,LesgourguesMangano2013}. For a
collisionless particle species with velocity $v(a)$, the comoving
free-streaming horizon is~\cite{Lesgourgues2006}
\begin{equation}
  \lambda_\mathrm{FS}(a) = \int_0^a \frac{da'\,v(a')}{a'^2 H(a')}.
  \label{eq:lambda_FS}
\end{equation}
For a Fermi--Dirac distribution the average momentum is $\langle p
\rangle \simeq 3.15\,T_s$~\cite{LesgourguesMangano2013}. Because $T_s
\propto a^{-1}$, the velocity redshifts as $v(a) \propto a^{-1}$ after
the non-relativistic transition. For $m_s \sim 1\,\mathrm{GeV}$ and
$T_\mathrm{eq} \simeq 0.8\,\mathrm{eV}$~\cite{Planck2018}, the velocity
dispersion at equality is $v_\mathrm{eq} \sim 10^{-9}$, so the
free-streaming scale is negligible compared to cosmological structure
scales. Consequently GeV-scale sterile neutrinos behave effectively as
cold matter with respect to structure
formation~\cite{Lesgourgues2006,LesgourguesMangano2013}. Their dominant
cosmological impact arises from their contribution to the background
expansion rate.

\subsection{Implications for CMB Observables}
\label{sec:cmb}

The epoch of matter--radiation equality leaves a characteristic imprint
on the CMB anisotropy spectrum through its influence on the early
integrated Sachs--Wolfe effect and the growth of
perturbations~\cite{Dodelson2003,Planck2018}. Planck constrains the
equality redshift with percent-level precision~\cite{Planck2018}. In
the regime relevant for GeV-scale sterile neutrinos, the equality shift
simplifies to $\Delta a_\mathrm{eq}/a_\mathrm{eq} \simeq -f_s$,
implying $f_s(a_\mathrm{eq}) \lesssim \mathcal{O}(10^{-2})$ from
consistency with current data. A complete assessment would require
incorporating the sterile component into \textsc{Class}~\cite{Lesgourgues2011}
or \textsc{Camb}~\cite{Lewis2000,Lewis2002} to evolve the perturbation
hierarchy. We leave this to future work.

\section{Numerical Validation of the Analytic Framework}
\label{sec:numerical}

To illustrate the behavior of the sterile neutrino component derived in
the previous sections, we perform a numerical evaluation of the
phase-space integrals for the energy density and pressure. The sterile
neutrino distribution is taken to be $f_s(q) = \alpha/(e^q+1)$, where
$q = ap$ is the comoving momentum normalized by the sterile temperature.
For definiteness we adopt $g_s = 2$ and consider representative values
of $\alpha$ within the partially thermalized regime. The sterile energy
density $\rho_s(a)$ is obtained by numerically integrating the
continuity equation~(\ref{eq:continuity}) with $w_s(a)$ computed
directly from the phase-space integrals
(\ref{eq:rho_s_q})--(\ref{eq:P_s_q}).

\subsection{Evolution of the Equation of State}

\begin{figure}[htbp]
  \centering
  \includegraphics[width=0.82\textwidth]{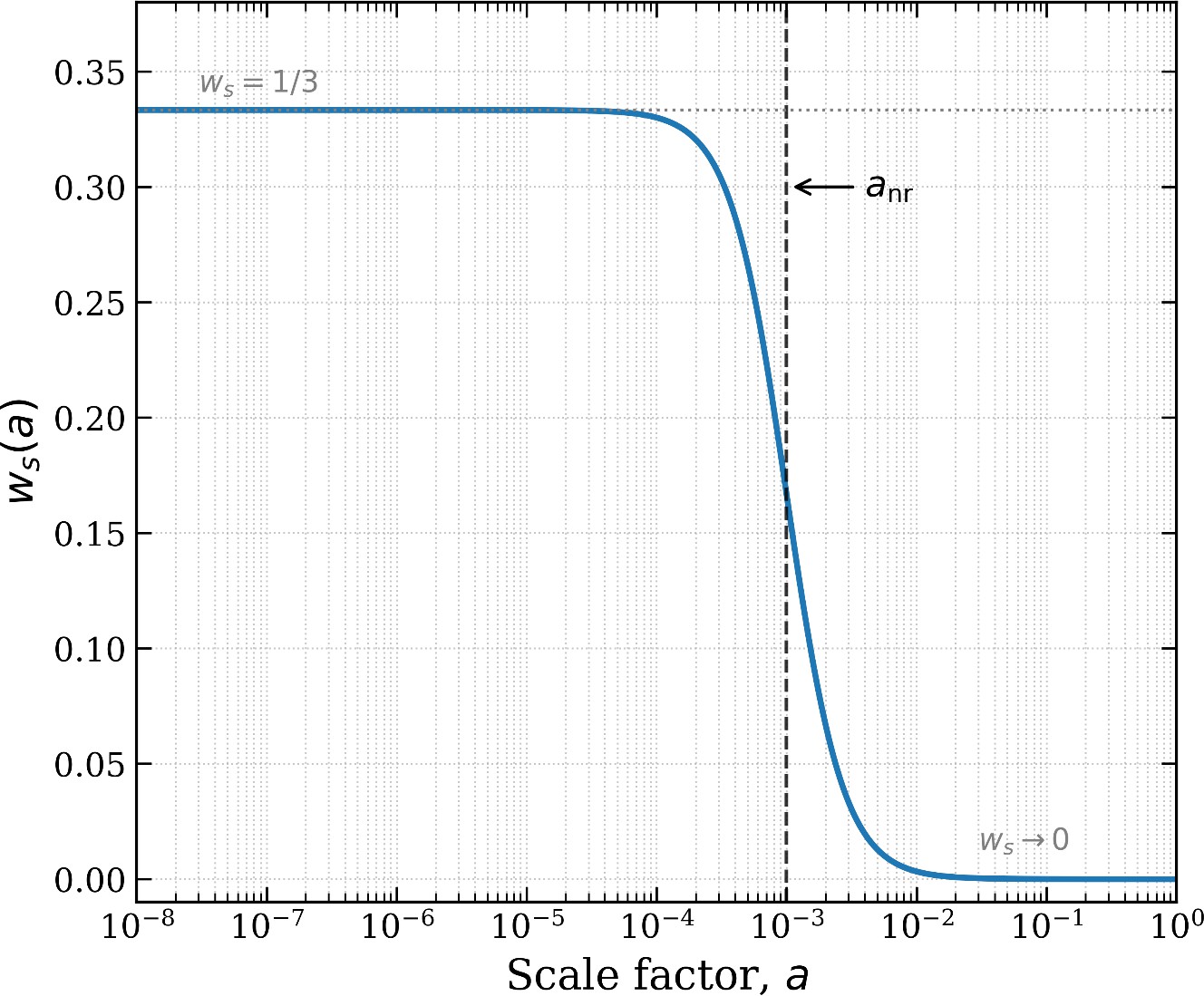}
  \caption{Evolution of the sterile neutrino equation-of-state
    parameter $w_s(a)$. At early times ($a \ll a_\mathrm{nr}$), the
    sterile population is relativistic and behaves as radiation with
    $w_s \approx 1/3$. As the Universe expands, the particles transition
    to non-relativistic behavior near $a \sim a_\mathrm{nr}$, leading to
    $w_s \to 0$. The vertical dashed line marks $a_\mathrm{nr} =
    10^{-3}$. This continuous evolution underlies the dynamical fluid
    description and cannot be captured by a fixed $\Delta N_\mathrm{eff}$
    offset.}
  \label{fig:eos}
\end{figure}

Using the expressions derived in Sec.~\ref{sec:rho_p} we compute the
equation-of-state parameter $w_s(a) = P_s(a)/\rho_s(a)$. The resulting
evolution is shown in Fig.~\ref{fig:eos}. At early times, when $T_s \gg
m_s$, the particles are relativistic and $w_s \to 1/3$~\cite{Lesgourgues2006,LesgourguesMangano2013}.
As the Universe expands and the temperature decreases, the sterile
population undergoes the relativistic--non-relativistic transition
(Regime~II in Sec.~\ref{sec:three_regimes}). During this period the
equation of state decreases smoothly toward $w_s \to 0$, corresponding
to non-relativistic matter. This transition is responsible for the
time-dependent behavior of the sterile energy density discussed in
Sec.~\ref{sec:friedmann}.

\subsection{Expansion-Rate Correction}

We next evaluate $\Delta H/H = (1/2)\rho_s/(\rho_r+\rho_m)$. The
numerical results confirm the analytic estimates derived in
Sec.~\ref{sec:param_estimate} and the three-regime structure described
in Sec.~\ref{sec:three_regimes}. During the relativistic phase the
sterile energy density scales as radiation, producing a nearly constant
fractional correction to the expansion rate consistent with the analytic
estimate $0.009\alpha$~\cite{Baumann2018,Kolb1990}. The largest
deviations occur near the relativistic--non-relativistic transition.

\subsection{Parameter Dependence}

The magnitude of the expansion-rate correction depends primarily on the
thermalization parameter $\alpha$. For representative values $\alpha
\sim 10^{-2}$--$10^{-1}$, corresponding to partially thermalized
sterile neutrinos consistent with the parameter window of
Fig.~\ref{fig:param_space}, the resulting corrections to the expansion
rate remain at the permille to percent level. This range is consistent
with the analytic estimates derived in Sec.~\ref{sec:param_estimate} and
with the observational constraints discussed in
Sec.~\ref{sec:constraint}.

\section{Limitations of the Present Framework}
\label{sec:limitations}

First, the sterile neutrino distribution was approximated by a
suppressed Fermi--Dirac form $f_s(p) \simeq \alpha
f_s^\mathrm{eq}(p)$~\cite{Dodelson1994,Dolgov2002}. This ansatz
captures the leading effect of incomplete thermalization but neglects
possible spectral distortions arising in detailed kinetic
calculations~\cite{Shi1999,Barbieri1991}. A more complete treatment
would involve solving the full Boltzmann
equation~\cite{Dodelson2003,Kolb1990} for the sterile neutrino
phase-space distribution.

Second, the present analysis focuses exclusively on the homogeneous
background evolution~\cite{Weinberg2008,Baumann2018}. A full assessment
of the impact on CMB anisotropies and large-scale structure requires
incorporating the sterile component into
\textsc{Class}~\cite{Lesgourgues2011} or
\textsc{Camb}~\cite{Lewis2000,Lewis2002}.

Third, we have restricted attention to a minimal scenario involving a
single sterile neutrino species with a thermal-like momentum
distribution. More general sterile neutrino sectors may involve multiple
sterile states or nonthermal production
mechanisms~\cite{Shi1999,Kusenko2009,Abazajian2017,Barbieri1991}.

Finally, our analytic estimates rely on order-of-magnitude constraints
derived from the stability of the matter--radiation equality
epoch~\cite{Planck2018}. A more precise determination would require a
global analysis including CMB, large-scale structure, and BBN
data~\cite{Planck2018,Lesgourgues2011,Lewis2000,Hannestad2010}.

\section{Discussion and Outlook}
\label{sec:discussion}

In this work we have developed an analytic framework for describing the
cosmological effects of partially thermalized sterile neutrinos with
finite mass. Rather than parameterizing the sterile contribution through
a constant shift in $\Delta
N_\mathrm{eff}$~\cite{Lesgourgues2014,Planck2018}, we treated the
sterile population as a dynamical cosmological fluid whose equation of
state evolves as the Universe expands~\cite{Weinberg2008,Dodelson2003}.

Starting from the Type~I seesaw Lagrangian (Sec.~\ref{sec:seesaw}), we
showed that incomplete thermalization arises naturally from the
suppression of the effective mixing angle in the primordial plasma
(Sec.~\ref{sec:micro}). The resulting suppressed Fermi--Dirac
distribution~\cite{Dodelson1994,Dolgov2002} was used to compute the
sterile energy density and pressure and incorporate them
self-consistently into the Friedmann
equations~\cite{Weinberg2008,Baumann2018}.

A central result is that partially thermalized sterile neutrinos
generically produce time-dependent modifications to the Hubble expansion
rate with three distinct dynamical regimes
(Sec.~\ref{sec:three_regimes}): a relativistic plateau at $\Delta H/H
\sim 0.009\alpha$, a relativistic--non-relativistic transition where the
dynamical fluid description most strongly departs from $\Delta
N_\mathrm{eff}$ approximations, and a matter-like growth phase where the
sterile component's relative contribution grows relative to
radiation~\cite{Lesgourgues2006,LesgourguesMangano2013}.

We applied this framework to the epoch of matter--radiation
equality~\cite{Dodelson2003,Planck2018}. For GeV-scale sterile
neutrinos the relativistic fraction at equality is negligibly small, so
the sterile population contributes as pressureless matter, leading to a
shift $\Delta a_\mathrm{eq}/a_\mathrm{eq} \simeq -f_s$. Consistency
with Planck observations~\cite{Planck2018} yields the approximate bound
$f_s(a_\mathrm{eq}) \lesssim \mathcal{O}(10^{-2})$
(Eq.~\ref{eq:fs_bound}). This constraint probes a phenomenological
regime that cannot be captured by the conventional $\Delta
N_\mathrm{eff}$
parameterization~\cite{Lesgourgues2014,Planck2018,Hannestad2010},
providing a complementary window on sterile neutrino cosmology.

We have also examined the free-streaming properties of GeV-scale
sterile neutrinos and found that their velocity dispersion at equality
is negligible, so they behave as cold matter with respect to structure
formation~\cite{Lesgourgues2006,LesgourguesMangano2013}.

Several extensions would be valuable. Incorporating the sterile
component into \textsc{Class}~\cite{Lesgourgues2011} or
\textsc{Camb}~\cite{Lewis2000,Lewis2002} would allow direct comparison
with CMB observations and refine the analytic bounds derived here.
Exploring nonthermal momentum distributions arising from resonant
production~\cite{Shi1999} or decays of heavier particles would extend
the framework to a broader class of production scenarios. In summary,
sterile neutrinos with incomplete thermalization can modify the
expansion history in a way not captured by the standard $\Delta
N_\mathrm{eff}$ description. The analytic approach developed here
provides a transparent starting point for exploring such effects and for
connecting sterile neutrino microphysics with observable cosmological
signatures.

\appendix
\section{Phenomenological Ansatz for the Sterile Neutrino Fluid}

In this work, the sterile neutrino component is modeled as an effective
cosmological fluid with a time-dependent equation-of-state parameter
$w_s(a)$. The goal of this construction is to capture the dominant
physical effect of the sterile sector---namely, its transition from
relativistic to non-relativistic behavior---without relying on a full
phase-space treatment.

The parameters $a_\mathrm{nr}$ and $\Delta$ can in principle be related
to the sterile mass scale and production mechanism, though we treat them
here phenomenologically.

At early times (small scale factor), the sterile population is
relativistic, and its pressure-to-density ratio approaches $w_s = 1/3$.
At late times, as the Universe expands and the typical momentum
redshifts below the mass scale, the species behaves as non-relativistic
matter with $w_s = 0$.

To describe this transition, we adopt a smooth interpolation between
these two limits. The transition is assumed to occur around a
characteristic scale factor $a_\mathrm{nr}$, corresponding to the epoch
at which the sterile species becomes non-relativistic. A second
parameter controls the width of the transition, ensuring that the
evolution is continuous and differentiable.

This parameterization satisfies the required asymptotic limits and
avoids unphysical discontinuities. It is therefore suitable for
incorporation into cosmological evolution equations where stability and
smoothness are essential. Such phenomenological ans\"{a}tze are commonly
employed in cosmology when the underlying microphysics is either
uncertain or not explicitly modeled, but the limiting behaviors are well
understood.

\section{Consistency with Energy Density Evolution}

Given a time-dependent equation-of-state parameter $w_s(a)$, the energy
density of the sterile component evolves according to the standard
continuity equation for a cosmological fluid, Eq.~(\ref{eq:continuity}).

This equation ensures that the evolution of the energy density is
consistent with the expansion of the Universe and the effective pressure
of the component. When integrated, it yields a solution in which the
energy density scales differently in the relativistic and
non-relativistic regimes.

In particular, at early times the energy density scales as $a^{-4}$,
consistent with radiation, while at late times it scales as $a^{-3}$,
as expected for non-relativistic matter. The smooth interpolation in
$w_s(a)$ ensures that the transition between these regimes is
continuous. This demonstrates that the adopted ansatz is not only
physically motivated but also dynamically consistent with standard
cosmological evolution.


\end{document}